\documentclass[runningheads]{llncs}
\usepackage{graphicx}
\usepackage{listings}
\usepackage[draft]{minted}

\usepackage{ dsfont }
\usepackage{ amsmath }
\usepackage{ eufrak }
\usepackage{enumitem}

\usepackage{todonotes}

\usepackage{lineno}

\begin{document}
\title{Towards Clause Learning \`a la Carte through VarMonads}
\titlerunning{Clause Learning \`a la Carte}
%
\author{Arved Friedemann\inst{1}\orcidID{0000-0001-7252-264X} \and\\
Oliver Keszocze\inst{2}\orcidID{0000-0003-2033-6153} }
\authorrunning{A. Friedemann, O. Keszöcze }
%
\institute{Swansea University, UK \\
\email{a.r.h.friedemann@swansea.ac.uk}\and
Friedrich-Alexander-Universität,
Erlangen-Nürnberg, Germany \\
\email{oliver.keszoecze@fau.de} 
}
%
\maketitle              
\begin{abstract}
More and more languages have a need for constraint solving capabilities for features like error detection or automatic code generation. Imagine a dependently typed language that can immediately implement a program as soon as its type is given. In SAT-solving, there have been several techniques to speed up a search process for satisfying assignments to variables that could be used for program synthesis. One of these techniques is clause learning where, if a search branch runs into a conflict, the cause of the conflict is analysed and used to create a new clause that lets a branch fail earlier if the conflict arises again. We provide a framework with which this technique can come for free not just for Boolean solvers, but for any constraint solver running on recursive algebraic data types. We achieve this by tracking the read operations that happen before a variable is assigned and use this information to create the dependency graph needed for conflict analysis. 

Our results are implemented in Agda for best readability, but they transfer to other functional languages as well. For brevity, we do not provide an entire search system utilizing the clause learning, but it will become clear from the formalisms that our technique indeed enables a clause learning search system to be built.



\end{abstract}
\section{Introduction}

Search is probably the most reoccurring problem in computer science, since any problem can be formulated as a search problem from the input to the desired output. One of the most general search problems is program synthesis, where we search for a general recursive program that fits a set of constraints or that has a certain type. In fact, in type theory, proof search and program synthesis are the same thing \cite{de1995curry,hottbook}. 
 If we could create a general search procedure for program synthesis, we would have a widely reusable tool for problem solving in computer science. To move towards a general solution to program synthesis we provide a type theoretical framework that makes search techniques from search or solving engines available to general programs. Specifically, we provide the basis to allow for clause learning in any search that can be described by a general recursive program. 

The high performance solving engines available to us thus far \cite{WCDHNR19} 
 are for primitive languages like Boolean satisfiability (SAT) \cite{biere2009handbook} or Sat-Modulo-Theories (SMT) \cite{Barrett2018}. The technique they use that is relevant for this paper is the Conflict-Driven-Clause-Learning, short clause learning, where an analysis of what caused a conflict is used to speed up the search process.

There have been functional approaches to solving, especially in the constraint solving field. A center piece of this development is \cite{schrijvers2009monadic}, where a highly adaptable framework for monadic constraint solving is implemented. This framework supports a version of constraint propagation (comparable to unit propagation used in SAT and SMT solving) and exchangeable search strategies through search nodes as first class objects. It features a variety of (mostly uninformed) search strategies such as (bounded) depth-first- or (bounded) breadth-first-search, but no clause learning. Because the approach in \cite{schrijvers2009monadic} is so highly adaptable, we will not present a complete solving system in this paper, but rather only the center piece that is needed for the clause learning. 

The approach in \cite{schrijvers2009monadic} has been extended to allow for an easier implementation of heuristics \cite{schrijvers2013search}, but clause learning is still left to SAT and SMT search engines \cite{uhler2014smten,armand2011modular,hallahan2019g2q}. These constraint satisfaction engines rely on variables that are filled with values to construct a model or to check consistency of constraints, which is the motivation for our paper. We use the fact that variables are the key component to make constraints communicate to implement a version of clause learning. We also call this variable-based constraint solving, which comes with a lot more benefits that will be compatible with our formalism. In \cite{kuper2013lvars,kuper2014freeze,newton2016parallel}, a system to make propagation via variables inherently concurrent was developed, so adding our principle to it even yields a possibility to do clause learning in parallel. What we add in this paper is a means to use the variables in any constraint solver to get clause learning for free. 

\subsection*{Contributions}

The main contribution of this paper is to create a structure for a free dependency graph for the evaluation of functions on recursive data types. In order to achieve this and to prove some correctness criteria like termination of the results, we implement the approach in Agda and provide an application of the Meta Theory \`a la Carte \cite{delaware2013meta} approach to datatypes that are distributed over memory via pointers.

\section{Preliminaries}
\subsection{Solving and Clause Learning}

We give a short summary of how solving engines work for SAT \cite{biere2009handbook} and hint at where these ideas generalise beyond the realm of Boolean satisfiability. 

In SAT, we are given a Boolean formula in Conjunctive Normal Form (CNF). Given a set of variables $\textit{VAR}$, let $C \subset\mathfrak{P}(\pm\textit{VAR})$, where both $C$ and $c\in C$ are finite. An element $pc_i\in c$ is a variable $c_i\in\textit{VAR}$ with a polarity $p\in\{-,+\}$ that indicates whether the variable in that clause is negated or not. The CNF has the form:

$$ \bigwedge_{c\in C} \bigvee_{pc_i\in c}pc_i $$
The search is for a satisfying assignment $\delta : \textit{VAR}\rightarrow \mathds{B}$.

The central solving part comes from the DPLL procedure \cite{DPLL} and is called the unit propagation. The idea is that if there is an unfullfilled clause with a single unsassigned literal $c_i$, the value of $c_i$ is assigned \textit{true} or \textit{false} depending on its polarity, so that the clause evaluates to \textit{true}. This assignment can cause other clauses to have only a single unassigned variable and the process continues. For each variable, we can memorise the reason why it was assigned. This reason is, that within the clause that triggered the assignment of the variable, all other variables have been assigned to the 'wrong' values. The representation of the reason is the list of variables and their assignments in the direct clause that led to the new assignment. This information is called the \emph{dependency graph}. If a conflict arises, meaning that a variable has been assigned conflicting values, it can be tracked what caused the conflict. This information is used for two things. First, it can deduce the earliest decision that led to the conflict, so the backtracking can go further back than it usually would, and second it is used to create a new clause that propagates the conflict earlier if it arises again in a different branch of the search. This last part is what's referred to as \emph{clause learning}.

When we generalise this idea, we don't need variables to only contain Boolean values, but they can contain any value that can be compared by equality (or values that can be merged until the merge causes a conflict, as in \cite{kuper2013lvars,kuper2014freeze,newton2016parallel}). The reason for an assignment is just the list of variables and their values that have been read until a new assignment was made. Obtaining this is what we refer to as \emph{dependency tracking}. A learned clause in the general setting is just a program that checks for the learned variable assignments and produces a conflict if necessary. 

\subsection{Agda and Dependent Type Theory}

The formalism for this paper uses literal Agda code. Agda is a functional language just like Haskell, but with a few additional features. In this paper, most of the difference are only of cosmetic nature, like mixfix operators (e.g. \mbox{\texttt{if\_then\_else\_}}), but we use three important concepts of dependent type theory.

The first concept is, that the type of a function can depend on its arguments. We mainly use this to give types as arguments to a function, e.g. written as \mbox{\texttt{(A : Set) -> List A}}, which is the type of a function taking a type and returning in this case a List that is parametrised by that type (note that Agda's type for types is called \texttt{Set}). If an argument to a function should be clear from the result, we write \mbox{\texttt{\{A : Set\} -> List A}}, where \texttt{A} is deduced based on how the function is used and \texttt{A} does not need to be provided explicitly, or \mbox{\texttt{\{\{Eq A\}\} -> List A -> Bool}} when the element of the type class \texttt{Eq A} should be uniquely deducible by search. These implicit arguments from the signature can, down in the function definition, be brought into scope by writing e.g. \texttt{\{A = B\}}, which takes the implicit variable \texttt{A} from the function signature and renames it to \texttt{B} (where quite often we don't want to rename the variable but just bring it into scope, so we just write \texttt{\{A = A\}}).

The second concept is, that we can apply functions on types within the type signature. This is the generalised version of a (dependent) type alias. So we can write \mbox{\texttt{Cumbersome A B = A -x- A -x- B -x- B}} (Where \texttt{-x-} is our ascii tuple type)

The third important concept is that of a sigma type. The type \texttt{Sigma A P} is a generalised version of a tuple where the first entry is an element \texttt{a} of \texttt{A} and the second entry is an element of \texttt{P a}, so \texttt{P} yields the type of the second entry. Imagine we have a pointer type \texttt{V}, where \texttt{V A} denotes a pointer that contains a value of \texttt{A}. Then, the tuple that contains a pointer and its value has the type \texttt{Sigma A V}. If we don't know the type of the pointer, we can have a sigma type that takes the type of the pointer as the first entry to the tuple, such as \verb$Sigma Set (\ A -> A -x- V A)$. A possible object of this tuple type could look like \texttt{(Bool, true, ptr 23)}.

We ignore one safety feature of Agda called universe-polymorphism. The idea is that \texttt{Set} itself needs a type, but if it were \texttt{Set : Set} then contradictions can happen. Agda solves this by writing \texttt{Set l : Set (suc l)}, but this results in additional boilerplate that we omit here for brevity.

Type classes in Agda are implemented via modules. Especially records (that can be automatically searched for by putting them into \texttt{\{\{...\}\}} brackets) serve as modules that can be opened with \texttt{open <record name> <record object>} to put their functions into the current scope. 

A few mentions of types and standard functions we are using. The string \texttt{-x-} was already mentioned to be the tuple type, with \texttt{\_,\_} being its constructor. \texttt{T} is the unit type with \texttt{tt} being its only, argumentless constructor. \texttt{\_\$\_} is the usual function application operator with lowest binding strength and \texttt{\_o\_} is function composition.   

\section{Motivating Example}\label{sec:Motivating_Example}

Let us look at the simple function 

\begin{minted}{agda}
any : List Bool -> Bool
any = foldr (_||_) false
\end{minted}

that computes whether at least one value in the list is true. When evaluating this function, we get an answer like

\begin{minted}{agda}
any [ false , true , false ] = true
\end{minted} 

We would now like to know why the function returned \texttt{true}. The reason is the part of the input that caused \texttt{any} to evaluate to its value. This could look like

\begin{minted}{agda}
any [ false , true , false ] = true because 
	any [ false , true , ... ] = true
\end{minted}

On this specific example, this is the smallest reason for the result. If the first value had been \texttt{true}, the reason would have been shorter. The second value was the final reason for the result being \texttt{true}. For search problems, these reasons are essential to guide the search process. Let us imagine a function

\begin{minted}{agda}
sudoku : (sfield : Mat 9 x 9 Nat) -> Bool
\end{minted}

which checks whether a given Sudoku-field filled with numbers is valid. To solve a sudoku, we might have to check several fields whether they are valid. Let's assume that we have a reason that 

\begin{minted}{agda}
sudoku sfield = false because 
	sfield at (2 , 3) = 5
	sfield at (5 , 3) = 5
\end{minted}

then, we no longer need to test fields that have a 5 at position (2,3) and (5,3), ruling out plenty of fields that no longer need to be branched on. We could retrieve these reasons by checking an entire (partial) field, but then returning only the portion of the fields that actually causes the field to be invalid, the same way as we did with the list in the first example. This is the idea of clause learning, generalised to a functional context. 

We mention here that this clause learning can become quite complex. The above example only provides the clause to a very direct conflict, but the most interesting conflicts would occur if there was one variable that cannot be assigned anything because of a sub-assignment. Imagine an assignment with 
\begin{align*}
\forall x,x'\in\{1,...,8\}.\ \texttt{sfield at (x , 9)}\neq\texttt{sfield at (x' , 9)}\\
\forall y,y'\in\{2,...,9\}.\ \texttt{sfield at (9 , y)}\neq\texttt{sfield at (9 , y')}
\end{align*}

Then the field (9,9) could not be assigned because it always violates one of the row or column constraints. If the search algorithm runs within the tracking of read variables, the above reason would be marked as the reason why the solver failed to assign (9,9) and therefore be giving an interesting clause. \\

Achieving this directly as shown is not possible in (pure) functional languages. This is because of the principle that the result of a function cannot depend on its evaluation. If a function could tell which values it read during evaluation, its results could change depending on how much of a value has been evaluated. There is however a way to achieve something similar in a functional language. Let us rewrite the any-function as follows

\begin{minted}{agda}
any lst = case lst of
	[] -> false
	(x :: xs) -> case x of
			true -> true
			false -> any xs
\end{minted}

Here we can see a bit more precisely how much of the value we actually read. Each case statement looks for the uppermost constructor to adjust its behavior. Assuming the code was written with as little case distinctions as necessary, this means that in each branch of the case-statement we know on which part of the input we currently depend on. Therefore, if our environment keeps track on which values we read on each case-statement, it is easy to acquire the dependency graph needed for clause learning. 

\section{VarMonads}

We will now define a type class that we can use to change the behaviour of the program depending on the variables that it read or wrote to, to add features like the creation of a dependency graph (this even allows for adding further features, such as instant parallelism or variable sharing...). This approach is inspired by a Haskell-Package called ``monad-var'' \cite{MonadVar} that offers type classes for pointer-like structures. 

In this paper we lift the monad-var approach to allow for simple clause learning and will discuss how it can be used to get even more pointer features for free. 

We start off with a type class that imitates Haskell's atomic IORef pointer model. In some monad $M$ with a pointer type $V$, we have three instructions:

\begin{minted}{agda}
record BaseVarMonad M V where
  field
	new : A -> M (V A)
	get : V A -> M A
	write : V A -> A -> M T
\end{minted}
\texttt{new a} creates a new pointer with the value \texttt{a}, \texttt{get p} returns the current value of the pointer \texttt{p} and \texttt{write p v} writes the given value into the pointer. This creates a basic structure for a VarMonad. 

The semantics of these VarMonads are generally as expected of a pointer interface, but the details can change based on the underlying monad. In \cite{kuper2013lvars,kuper2014freeze,newton2016parallel} e.g., the information content in a pointer can only ever increase, which is actually an important constraint needed for solving that we will handle later. We will use the VarMonads to track the \texttt{get} operations of a program to retrieve the reasons for assignments. Before we go into constructions on VarMonads, we will show how datatypes can be encoded in a way that can be handeled well in this pointer setting.

\section{Data Types and Meta Theory à la Carte}

Data types à la Carte (DTC) \cite{swierstra2008data} is an approach for having composable data structures. It was first introduced for Haskell and later ported to functional proof assistants like Coq and Agda \cite{delaware2013meta,forster2020coq,keuchel2013generic}, now called Meta Theory à la Carte (MTC) \cite{delaware2013meta}. The key idea is the following:

A data type is represented by a functor that takes the type of its ``recursive call". E.g., if we look at the following definition of a list:

\begin{minted}{agda}
data ListF (A : Set) : (B : Set) -> Set where
  nil : ListF A B
  lcons : A -> B -> ListF A B
\end{minted}

We can see that at the position where there would usually be the recursive call to \texttt{ListF A}, there is a \texttt{B} instead. This way, the recursive call can be exchanged to e.g. be a pointer to the sublist, instead of the sublist itself. This approach can be used for other features like data type composability, but here we only need the possibility to switch the recursive call to being a pointer. We now first explore how to use the above functor to create a list in the classical sense, and how to define a function over it.

If we want to have a list in the classical sense we need to have an operator that gives this list-functor itself as an argument to the recursive call. Naively, this type would be \mbox{\texttt{ListF A (ListF A (ListF A (...)))}}. In Haskell this is achieved with the following operator

\begin{minted}{haskell}
data Fix f = In f (Fix f)
\end{minted}

The type of A-lists can be obtained as \texttt{Fix (ListF A)}, which is constructed using the \texttt{In} constructor together with the list functor constructors that take another \texttt{Fix (ListF A)} as the recursive argument. The problem is that this \texttt{Fix} operator could potentially be used to express infinite data structures and is not allowed in languages like Agda or Coq. To fix this, a different, non-recursive operator is used that creates a church-encoding of the data type, effectively freezing the constructors with the function that reduces them during a fold (also called an ``algebra"). These are the operators used in Meta Theory à la Carte (MTC) \cite{delaware2013meta} (and have later been refined \cite{keuchel2018reusability,keuchel2013generic}, but we will stick with the original definitions here for simplicity). In this paper, we will use both approaches, but will for now focus most on the MTC approach. Using the approach provides additional guarantees like termination of our constructions. It also makes it possible for future research to have the correctness of the constructions verified. \\
The first Idea of MTC looks as follows:

\begin{minted}{agda}
Algebra : (F : Set -> Set) -> (A : Set) -> Set
Algebra F A = F A -> A

Fix : (F : Set -> Set) -> Set
Fix F = forall {A} -> Algebra F A -> A

foldF : Algebra F A -> Fix F -> A
foldF alg fa = fa alg
\end{minted}

The fixpoint of a functor is now a function that takes an arbitrary algebra to reduce the value to some type \texttt{A}. This freezes the constructors of that type until the function evaluating it is given. If we now construct the fixpoint of one of our data-functors, we can fold over it via \texttt{foldF}. The usage of this is best shown by the example of how to create the any-function for our list-functor:


\begin{minted}{agda}
anyFL : Fix (ListF Bool) -> Bool
anyFL = foldF \ {
  nil -> false;
  (lcons x xs) -> x || xs}
\end{minted}

The important part for our VarMonads is now: Instead of the direct recursive call, we can put a pointer to the recursive substructure. That way, even arbitrarily large data structures are inherently distributed over several variables (for each of which we can do the dependency tracking separately). With a structure distributed over a state with pointers, a naive implementation of the any-function could look like

\begin{minted}{agda}
anyVM : {{bvm : BaseVarMonad M V}} -> Fix (ListF Bool o V) -> M Bool
anyVM = foldF \ {
    nil -> return false;
    (lcons x xs) -> (x ||_) <$> (join $ get xs)}
\end{minted}

If we would directly apply the clause learning to this however, this function would now always read the entire list and there would be no information on which parts had been important. To solve this issue, we use the Mendler-Style algebras as proposed in \cite{delaware2013meta,uustalu2000coding}. These algebras make explicit when the recusive value is used. 

\begin{minted}{agda}
Algebra : (F : Set -> Set) -> (A : Set) -> Set
Algebra F A = forall R -> ([[_]] : R -> A) -> F R -> A
\end{minted}

This is the algebra that we will use for the fixpoints in the context of VarMonads. The way that this works is that the functor \texttt{F} contains a value of an unknown type \texttt{R} that can only be extracted when the extra function given is applied (the brackets \texttt{[[ ]]}). In practise, this looks as follows. Let's say we want to write the any-function with the Mendler-style algebras. They look just like before, only that we wrap the \texttt{[[ ]]}-call around the result from the recursive call. 

\begin{minted}{agda}
anyFL : Fix (ListF Bool) -> Bool
anyFL = foldF \ {
  _ [[_]] nil -> false;
  _ [[_]] (lcons x xs) -> x || [[ xs ]]}
\end{minted}

This way, we can get information about whether the recursive call is used or not without having to rely on information from the lazyness. Our any-function now looks like:

\begin{minted}{agda}
anyDTC : {{bvm : BaseVarMonad M V}} -> Fix (ListF Bool o V) -> M Bool
anyDTC = foldF \ {
    _ [[_]] nil -> return false;
    _ [[_]] (lcons true xs) -> return true;
    _ [[_]] (lcons false xs) -> get xs >>= [[_]] }
\end{minted}

The \texttt{[[\_]]} function is only ever applied to the recursive call in the last case where we have to look at the rest of the list, so this function will only read exactly what is necessary to retrieve the information of the dependency graph and there are no side effects to the state when the recursive value is not used by the function. 

We can now almost ignore the explicit pointers themselves completely by formulating a special fold

\begin{minted}{agda}
foldBVM :
  {{bvm : BaseVarMonad M V}} ->
  Algebra F (M A) -> Fix (F o V) -> M A
foldBVM {{bvm}} alg = foldF \ _ [[_]] -> alg _ (get >=> [[_]])
\end{minted}

Here, we give an algebra that just acts on the original functor \texttt{F} whose recursive calls are fed with the result from the \texttt{get} action. The only downside is that the result \texttt{A} needs to be wrapped into the monadic context because retrieving the recursive value has a side effect within the monad. We can now write the any-function as

\begin{minted}{agda}
anyDTC : {{bvm : BaseVarMonad M V}} -> Fix (ListF Bool o V) -> M Bool
anyDTC = foldBVM {F = ListF Bool} \ {
    _ [[_]] nil -> return false;
    _ [[_]] (lcons true xs) -> return true;
    _ [[_]] (lcons false xs) -> [[ xs ]] }
\end{minted}

Where the \texttt{[[ ]]} brackets retrieve the result of the recursive call in a monadic context.

There is one more upside to using Mendler-style algebras. When constructing a fix-value, we no longer rely on a functor instance. Instead we can just write

\begin{minted}{agda}
In : F (Fix F) -> Fix F
In f A alg = alg _ (foldF alg) f
\end{minted}

However, we cannot get rid of the functor instance when extracting values. This looks as 

\begin{minted}{agda}
Ex : {{Functor F}} -> Fix F -> F (Fix F)
Ex = foldF \ _ [[_]] f -> In o [[_]] <$> f
\end{minted}
So, when we want to look at the uppermost constructor (for a show instance or merging operations or something) we need a functor instance. 

Pointers cannot have a functor instance for the simple fact they have to be writeable, which would mean that we could also write any value of the new type after the map. This cannot be done as not all functions used in a map are surjective. With the \mintinline{agda}{Ex}-function however, we have a special case where the type nor value of the pointer are actually changed. As we can safely assume all \mintinline{agda}{Fix}-values were created with \mintinline{agda}{In}, the \mintinline{agda}{Ex}-function recurses over \mintinline{agda}{Fix}-values that have been extracted with \mintinline{agda}{Ex} and are being packed back in using \mintinline{agda}{In}, which is just like applying \mintinline{agda}{id}.

Therefore, it is fine to create a read-only variant for pointers that has a functor instance for the \mintinline{agda}{Ex}-function. We call our variant ``Lensed Variables" (which is related to but not the same as general functional lenses).

A lensed variable has an original variable and type stored, but also additionally a function that can transform the original value into any other value. This way, two parts of the program can look at the same variable through different ``lenses":

\begin{minted}{agda}
data LensPtr (V : Set -> Set) (A : Set) where
	constructor LP
  field
	origType : Set
	origPtr : V origType
	t : origType -> A
\end{minted}
We can give such a Pointer a Functor instance as follows:

\begin{minted}{agda}
LensPtrFunctor : Functor (LensPtr V)
LensPtrFunctor = record { _<$>_ = \ f (LP T p t) -> LP T p (f o t)}
\end{minted}
Where the pointer has an additional transformation to its output. Now we can use the normal MTC constructions (like \texttt{Ex}) that require functor instances with pointer structures as long as we only apply the \mintinline{agda}{id} functor or only read from the pointer. 

\section{Variable Tracking}\label{sec:Variable_Tracking}

Now that we have lifted all functions into our VarMonad, we want to track the values that have been read. This can be achieved with a simple state transformer. Before that, we first settle on the type of data that will be tracked.

\begin{minted}{agda}
AsmCont : (C : Set -> Set) -> (V : Set -> Set) -> Set
AsmCont C V = C (Sigma Set \A -> (A -x- V A))
\end{minted}

We define an assignment-container \texttt{AsmCont} to be a functor that contains tuple variables with the value they had when read. The \texttt{Sigma} type is a dependent tuple, in this case meaning that we actually store a triple with first the type of the pointer value, then the value and then the pointer itself. We define a special VarMonad, a \texttt{TrackVarMonad} that has a BaseVarMonad \texttt{bvm} and an additional action \texttt{getCurrAssignments} to retrieve the currently read values.
\begin{minted}{agda}
record TrackVarMonad ...
    getCurrAssignments : M (AsmCont C V)
\end{minted}

We can create an instance of this monad via the following construction where we add the read values to the state. Here we assume the containter \texttt{C} to have a \texttt{MonadPlus} instance, so that they can behave like sets with a merge and a singleton operation. 

\begin{minted}{agda}
BaseVarMonad=>TrackVarMonad : {{mpc : MonadPlus C}} ->
  BaseVarMonad M V ->
  TrackVarMonad C (StateT (AsmCont C V) M) V
BaseVarMonad=>TrackVarMonad {C = C} bbvm = record {
    bvm = record {
      new = liftT o new ;
      get = \ {A = A} p -> do
        v <- liftT (get p)
        modifyS (_<|> return (A , v , p))
        return v
        ;
      write = \ p -> liftT o write p } ;
    getCurrAssignments = getS }
  where
    open BaseVarMonad bbvm
\end{minted}

This construction is straightforward, as it accumulates the read values into the state. A welcome side effect is that now, through the alternative-instance of the state monad, parallelism or different programs reading different variables comes for free (though we do not use it in the scope of this paper).

\section{Product VarMonads}

Ultimately, we want to write the list of reads into every variable. We want to hide this extra information behind a VarMonad that can only retrieve the extra value, so we define a new specialised VarMonad
\begin{minted}{agda}
record SpecVarMonad ...
    get : V A -> M B
    write : V A -> B -> M T
\end{minted}

This VarMonad has an extra \texttt{get} and \texttt{write} operation for the extra attached value of every pointer. A na\"ive implementation to create the tuple of a corresponding BaseVarMonad and SpecVarMonad pair would be a construction

\begin{minted}{agda}
naiveProd : BaseVarMonad M V -> (B : Set) -> 
	    	BaseVarMonad M (\ A -> V (A -x- B))
		-x- SpecVarMonad M (\ A -> V (A -x- B))
\end{minted}

This constructs a pair of VarMonads that only operate on pointers of tuples with the main value and the additional value. This however is not expressive enough. If we look at the type of the reasons from Section \ref{sec:Variable_Tracking}, we notice that the type B depends on the general variable type of the monad. This is similar to having a pointer that eventually points to itself, like a type \mbox{\texttt{V (A -x- V (A -x-...))}}. This means that we need a pointer type

\begin{minted}{agda}
FullAsmPtr M V C A = 
	V (A -x- Fix (\ R -> AsmCont C (\ B -> V (B -x- R)) ) )
\end{minted}

where the \texttt{AsmCont} container stores the tuple based variables. We generalise this type for our product construction
\begin{minted}{agda}
RecTupPtr M V F A = V ( A -x- Fix (\R -> F (\ B -> V (B -x- R) ) ) )
\end{minted}

where we allow a general container \texttt{F} to take an arbitrary pointer type that is exchanged with the recursive tuple pointers. Our assignment-container pointer will later look like

\begin{minted}{agda}
AsmPtr M V C = RecTupPtr M V (AsmCont C)
\end{minted}
%
We now formulate the product construction as

\begin{minted}{agda}
recProdVarMonad : BaseVarMonad M V -> 
	{B : Set} -> {F : (Set -> Set) -> Set} ->
	{{func : Functor (\ R -> F (\B -> V (B -x- R)))}} ->
	(forall {V'} -> F V') ->
	    BaseVarMonad M (RecTupPtr M V F) 
	-x- SpecVarMonad M (RecTupPtr M V F) (F (RecTupPtr M V F))
recProdVarMonad bvm mpty = (record {
      new = new o (_, In mpty) ;
      get = (fst <$>_) o get ;
      write = \ p v -> snd <$> get p >>= \b -> write p (v , b) }
    ) , (record {
      get = \ p -> snd <$> get p >>= Ex ;
      write = \ p v -> fst <$> get p >>= \ a -> write p (a , In v) })
  where open BaseVarMonad bvm
\end{minted}
First, we need the container \texttt{F} to be a Functor, at least for tuple based pointers, so that it can be deconstructed via \texttt{Ex}. Then we need it to have an empty element \texttt{mpty}, so that it can be created alongside a value without further information. The \texttt{BaseVarMonad} is now created in a straightforward manner, where the original value is just wrapped into the tuple and if one value of the tuple is updated, the other one is read and untouched.

We notice how, in this construction, we always override the old values with either the variables or the reasons. For a correctly working system, old values have to be preserved somehow (e.g. by using lattices for merging values), but we omit that detail here for brevity.

 The \texttt{SpecVarMonad} now acts on the second element of the tuple in each pointer. It is being implemented analogously otherwise. We note that as soon as one constrains all values to be (bounded) lattices, the boilerplate for reading old values and constructing default values can be removed. This however makes the type signatures more complicated than currently necessary.

\section{Dependency Graph}

We now combine the previous two constructions to track the reasons for every assignment and writing it into the values themselves. We create a Clause-Learning VarMonad for the purpose. It, again, has a BaseVarMonad \texttt{bvm} and has a special operation \texttt{getReasons}
\begin{minted}{agda}
record CLVarMonad ...
    getReasons : V A -> M $ C (AsmCont C V)
\end{minted}
This VarMonad has an extra operation to retrieve the reasons from a pointer. There can be several reasons, because the pointer might be assigned several times. The construction for this VarMonad looks as follows:

\begin{minted}{agda}
BaseVarMonad=>CLVarMonad : BaseVarMonad M V ->
  (forall {A} -> C A) ->
  {{func : Functor (\ R -> C $ AsmCont C (\B -> V (B -x- R)))}} ->
  {{mplus : MonadPlus C}} ->
  CLVarMonad (StateT (AsmCont C (AsmPtr M V C)) M) (AsmPtr M V C) C
BaseVarMonad=>CLVarMonad {M} {V = V} {C = C} bvm mpty = record {
    bvm = record {
      new = \ x -> new x >>= putAssignments ;
      get = get ;
      write = \ p v -> putAssignments p >> write p v };
    getReasons = getR }
  where
    vmtup = recProdVarMonad bvm 
    	{B = C $ AsmCont C (AsmPtr M V C)} 
    	{F = C o AsmCont C} mpty
    trackM = BaseVarMonad=>TrackVarMonad (fst vmtup)
    lspec = liftSpecVarMonad (snd vmtup)
    open BaseVarMonad bvm using (mon)
    open TrackVarMonad trackM
    open SpecVarMonad lspec renaming (get to getR; write to writeR)
    putAssignments : AsmPtr M V C A -> 
    	StateT (AsmCont C (AsmPtr M V C)) M (AsmPtr M V C A)
    putAssignments p = 
    	getCurrAssignments >>= writeR p o return >> return p
\end{minted}
The important parts are the constructions around the \texttt{bvm}. First, we apply the product construction to create pointers that hold the reasons as well. We give the type of pointer content explicitly to make it easier for Agda to figure out the types. Then, we apply the tracking to just the BaseVarMonad from the product, using the SpecVarMonad to retrieve the reasons from a pointer. The SpecVarMonad needs to be lifted to still work within the StateT transformer. We use the TackVarMonad as our main VarMonad and then just put the current remembered assignments into the modified pointers. 

With this construction, we now get a VarMonad that can tell for every value why it has been assigned (if we only perform minimal reads). 

\section{Constrained Values}

One last feature that should be mentioned is that, with a bit of extra code, the values used in the VarMonad can be constraint to all have certain properties. This can be useful to give all values additional features like a \texttt{Show}, \texttt{Eq} or even a \texttt{Lattice} instance. As the interface of the \texttt{BaseVarMonad} changes, the above constructions have to be redone, but we decided to give the raw definitions first to abstract from unnecessary detail. This is the definition of a constrained VarMonad, that takes an extra parameter \texttt{K} that serves as a predicate on the values within the pointers.
\begin{minted}{agda}
record ConstrVarMonad ...
    new : {{k : K A}} -> A -> M (V A)
    get : {{k : K A}} -> V A -> M A
    write : {{k : K A}} -> V A -> A -> M T
\end{minted}
Let e.g. \texttt{K} = \texttt{Show}, and all values stored in variables can be converted into text. There can also be several constraints, e.g. by choosing 

\verb$K = \ A -> Show A -x- Eq A -x- Lattice A $\\
The constructions only change marginally, however, some constructions become simpler with certain constraints. If \texttt{K} implies a (bounded) lattice, the \texttt{write} function can be implemented as 

\begin{minted}{agda}
write p v = get p >>= \v' -> write p (v /\ v')
\end{minted}

And then the product construction does not need to handle retrieving the old value (or constructing a default value).

We even note that the lattice constraint is essential for the constructions to have the desired semantics. If a variable can be arbitrarily reassigned, any bookkeeping of its value gets invalidated. This does not happen when the value can only grow in information content. 

In this paper, we are only using these constraints for show instances, but e.g. for the recursive construction of an overall reason (by traversing the dependency graph), all values need to have an \texttt{Eq} instance to ensure termination. 

\section{Running the Constructions}

We run our constructions with a default implementation. The default BaseVarMonad runs on a state with a map, mapping integers to values. From this BaseVarMonad, we derive a CLVarMonad with our constructions and let the initial list example run. We adapted the list constructors in a way so that they directly create a list with a pointer at the recursive call.

\begin{minted}{agda}
anyTest : Bool
anyTest = runDefCLVarMonad $ do
  anyDTC =<< false :: true :: false :: []
\end{minted}

Note how on the top level, we use the DTC implementation of \texttt{any} in order to only read necessary values. 

For simplicity, we omit the \texttt{runDefCLVarMonad} boilerplate from now on. We can now extract the reasons as follows:
\begin{minted}{agda}
do
  res <- new =<< anyDTC =<< false :: true :: false :: []
  getReasons res
\end{minted}
Together with the show constraints, we get the list of reasons

\begin{minted}{agda}
"(p3 = lcons false p2) ^ (p2 = lcons true p1)"
\end{minted}
Where \texttt{p<i>} are the pointers in which the values are stored in. This is precisely the result that we wanted. As we have used the DTC approach, this now works for all functions on DTC data types.

\section{Summary and Future Work}

In this paper, we have created a mechanism to retrieve the reason as to why a value has been assigned. This approach works with all DTC and MTC data types and can be used in a solving system to implement clause learning. The underlying implementation was created using Agda, giving safety features like type safety and a guarantee of termination. Only checking consistency with universe polymorphism is left to future work. We lifted the MTC approach to work within our monadic structures and showed that all DTC (and MTC) data types and functions over them can be represented in our system. As our constructions use a state monad for the current assignment and dependency tracking, we get search features like branching for free. Therefore, for future work, our approach only needs to be combined with \cite{schrijvers2009monadic} to create a full solving system. Furthermore, it needs to be investigated how to combine the results with \cite{kuper2013lvars,kuper2014freeze,newton2016parallel} so that the constraint solving and clause learning come for free even in a concurrent setting. For now, we have built the center piece for clause learning and we are one step closer to building a general purpose, functional solving system.

%
%
%
\bibliographystyle{splncs04}
\bibliography{base}

\end{document}